\documentclass[secnumarabic,aps,prd,floatfix,nofootinbib,showkeys,twocolumn,showpacs,10pt,superscriptaddress]
{revtex4-1}
\usepackage{times}
\usepackage{amsfonts}
\usepackage{amsmath}
\usepackage{amssymb}
\usepackage{natbib}
\usepackage{graphicx}
\usepackage{xcolor}
\usepackage[colorlinks=true,linkcolor=blue]{hyperref}
\usepackage[outdir=./]{epstopdf}
\usepackage[normalem]{ulem}
\usepackage{enumerate}
\usepackage[inline]{enumitem}
\usepackage{moreenum}

\def\aap{A\&A}
\def\aj{AJ}
\def\apj{ApJ}
\def\apjl{ApJL}
\def\apjs{ApJS}
\def\apss{Astrophysics and Space Science}
\def\jcap{JCAP}
\def\mnras{MNRAS}

\def\pasp{PASP}

\begin{document}

\title{{Barions and 
\texorpdfstring{$\Lambda$CDM}{LCDM} model problems}
}

\author{A.~\surname{Del Popolo}}%
\affiliation{%
Dipartimento di Fisica e Astronomia, University Of Catania, Viale Andrea Doria 6, 95125, Catania, Italy
}

\affiliation{
Institute of Astronomy, Russian Academy of Sciences, 119017, Pyatnitskaya str., 48, Moscow
}

\email{adelpopolo@oact.inaf.it}

\author{Morgan~\surname{Le~Delliou}}
\affiliation{Institute of Theoretical Physics, School of Physical Science and Technology, Lanzhou University, No.222, South Tianshui Road, Lanzhou, Gansu 730000, China}
\affiliation{Instituto de Astrof\'isica e Ci\^encias do Espa\c co, Universidade de Lisboa, Faculdade de Ci\^encias, Ed. C8, Campo Grande, 1769-016 Lisboa, Portugal}
\email[Corresponding author: ]{(delliou@lzu.edu.cn,) Morgan.LeDelliou.ift@gmail.com}
\affiliation{Lanzhou Center for Theoretical Physics, Key Laboratory of Theoretical Physics of Gansu Province, Lanzhou University, Lanzhou, Gansu 730000, China}

\label{firstpage}

\date{\today}

\begin{abstract}
We examine the inner profiles of the rotation curves of galaxies in the velocity range from dwarf galaxies to Milky-way like galaxies, having maximum of rotation, $V_{\rm max}$, in the range [30-250] km/s, whose diversity is much larger than that predicted by the $\Lambda$CDM model.
After showing that the scatter in the observed rotation curves is much larger that that predicted by dark matter-only cosmologies, we show how taking into account baryons, through a semi-analytical code, allows to create a a variety of rotation curves in agreement with observation. Simultaneoulsy, we show that the the quoted discrepancy does not need for different form of dark matter as advocated by \citep{Creasey2017}, and \cite{Kaplinghat2016}. 
We finally show how our model can reobtain the rotation curve of remarkable outliers like IC~2574, a 8 kpc cored profile having a challenging, and extremely low rising rotation curve, and UGC~5721 a cusp-like rotation curve galaxy.
We suggest treating baryonic physics properly before introducing new exotic features, albeit legitimate, in the standard cosmological model
\end{abstract}



\maketitle

\section{Introduction}
While the ``Concordance'', $\Lambda$CDM, model explains the main cosmological observations  \citep{Spergel2003,Komatsu2011}, it is not exempt of issues \citep[e.g.][]{Weinberg1989}. One particular such issue resides in the inconsistency \citep[e.g.][]{Navarro1996} between the dissipationless N-body simulations predictions of a cuspy density profile for structures, and the observed flatness of central profiles in dark-matter (DM) dominated  \begin{enumerate*}[label=\itshape\arabic*.\upshape)]                                                                                                                                                                                                                                    \item high surface brightness spiral galaxies 
\item Irregulars and Low Surface Brightness galaxies (hereafter LSBs)                                                                                                                                                                                                                                                                                                                                                                                                                                    \item  dwarf galaxies                                                                                                                                                                                                                                                          \end{enumerate*}
found in some galaxy clusters. This inconsistency was coined cusp/core problem \citep{Moore1994,Flores1994}.
This issue can be related to the Too-Big-To-Fail (TBTF) problem \citep{BoylanKolchin2011,Papastergis2015} and better interpreted as an excess of DM in the galaxies central regions.
Another such issue worth pointing out consists in the discrepancy between the poor diversity, in numerical simulations, and the rich diversity (hereafter dubbed ``diversity problem") in observations of dwarf-galaxies rotation curves (RCs).

A radical direction proposed to solve these issues, the rejection of the $\Lambda$CDM assumption that DM is cold, leads to a flurry of models \cite[e.g. self-interacting DM (SIDM) models][]{Spergel2000}. The latter model was recently included in a proposed solution to the $\Lambda$CDM model small scale problems \cite{Kaplinghat2016}.

In this proposition, several issues of the $\Lambda$CDM model and , in particular the rotation curve of IC~2574, are claimed to have almost no solutions within the ``Concordance'' model. Based on that claim, \cite{Kaplinghat2016} argue that a solution to those issues lie in the alternative SIDM model.

Testing whether the \cite{Kaplinghat2016} claim can be overturned within the $\Lambda$CDM model constitutes therefore a task of fundamental importance.

Such is the goal of the present paper. It will focus on the possibility,  without new physics, to solve the $\Lambda$CDM model's ``deficit problem in halos'' and will show how accounting for baryonic physics provides such solution.

The structure of the paper organises as follows: Sec.~\ref{sect:model} summarises our semi-analytical model, which allows more clearly and easily than N-body and hydrodynamical simulations to disentangle each physical effects. The model's solution to the discussed \cite{Oman2015} diversity problem is shown in Sec.~\ref{sect:diversity}. Sec.~\ref{sect:dwarf} explains with the model some peculiar galaxies RCs (e.g., IC~2574, UGC~5721). Note that the latter (UGC~5721) was previously fitted by a combination of SIDM and baryon physics \cite{Creasey2017}. Finally, we compare our results with \cite{Kaplinghat2016}'s and conclude in Sec.~\ref{sect:conclusions}

\section{Summary of our semi-analytical model}
\label{sect:model}
 This section proposes to summarise the model explained in \cite{DelPopolo2009,DelPopolo2016a,DelPopolo2016b}.
            {A
n} improvement to the spherical infall models (SIM) \cite{Gunn1972,Ryden1987,Williams2004}, {
  {it} 
includes} ordered \cite{Ryden1988} and random angular momentum \cite{Ryden1987,Williams2004}, adiabatic contraction 
\cite{Blumenthal1986,Gnedin2004}, the effect of dynamical friction of gas and stellar clumps on the DM halo 
\cite{ElZant2001,ElZant2004,Cole2011,Inoue2011,Nipoti2015}, gas cooling, star formation, photoionization, supernova, 
and AGN {(Active Galactic Nucleus)} feedback \cite{DeLucia2008,Li2010}.

{The model considers an initial proto-structure in the linear phase containing DM and baryons (diffuse gas) whose 
abundance is given by the ``universal baryon fraction" $f_{\rm b} = 0.167 \pm 0.01$ \citep{Komatsu2011}. Initial 
conditions are set as in \cite{DelPopolo2009}. This perturbation expands with the Hubble flow to a maximum radius 
(turn-around) and then
its dark matter re-collapses, and creates the potential well baryons 
will fall in. The DM and baryons equations of motion 
depend on the angular momentum (AM) acquired through tidal 
torques (ordered AM, $L$) \cite{Hoyle1953,Peebles1969,White1984,Ryden1988}, on the random AM, $j$, 
from collapse generated random velocities \cite{Ryden1987}, on the dynamical friction (DF) and on the 
cosmological constant. The 
ordered AM follows the standard Tidal-Torque-Theory (TTT), 
while the random AM, is assigned at turnaround, as described in \cite{DelPopolo2009}. The dynamical friction force 
between DM and baryonic clumps 
is evaluated as in \cite{Kandrup1980} and 
extends \cite{Chandrasekhar1943}. Clumps dimensions, lifetime and characteristics are described in 
\cite{DelPopolo2016b}. For spiral galaxies, they are obtained using Toomre's criterion \cite[see][]{DelPopolo2016b}. 
Clumps density and rotation velocity are similar to those found by 
\cite[see their Figs. 15 and 16]{Ceverino2012}.

DM 
adiabatically contracts 
during 
collapse 
and forms a more cuspy profile 
\citep{Blumenthal1986,Gnedin2004}. Assuming proportional 
DM and baryons initial density profiles 
\cite[i.e., choosing an NFW profile]{Keeton2001,Treu2002,Cardone2005}, $M_{\rm dm}$ is 
solved iteratively 
from 
angular momentum conservation 
\citep{Blumenthal1986,Ryden1988}.

Radiative processes cool down baryons that form clumps which collapse to the halo centre because of DF, while forming stars.
These clumps then transfer energy and AM to DM 
\citep{ElZant2001,ElZant2004}, increase their random motion and produce a predominant outward motion for DM particles, 
reducing the central density; the density cusp is heated and a core forms. This provides the main mechanism of core 
formation for dwarf spheroidals and spirals while for spiral galaxies, this mechanism is amplified by the acquired 
ordered and random AM.

A similar mechanism contributing to the core formation is given by SN feedback \citep{Pontzen2014}, in which repeated 
SN explosions flatten the profile. They are compared in detail in \cite{DelPopolo2016a}. 
Gas cooling is treated as a classical cooling flow \citep[see][]{Li2010}; star formation, reionization and SN feedback 
are included as in \cite{DeLucia2008,Li2010}. 
The star formation rate $\psi=\alpha_{\rm SF} M_{\rm sf}/t_{\rm dyn}$ depends on the star formation efficiency 
$\alpha_{\rm SF}=0.03$ \cite{Li2010}, on the dynamical time $t_{\rm dyn}$ and on the gas mass above a given density 
threshold $M_{\rm sf}$, fixed to $n>9.3~{\rm cm}^{-3}$ as in \cite{DiCintio2014}. The initial mass function (IMF) is 
chosen as Chabrier \cite{Chabrier2003}. At each time step $\Delta t$, $\Delta M_{\ast}=\psi\Delta t$ stars are 
generated \cite[see][for more details]{Li2010}.

The specific implementation for SN feedback is given in \cite{Croton2006} and {develop} as follows: the energy injected in the 
interstellar medium (ISM) by a SN explosion, $\Delta E_{\rm SN}$ is proportional \citep[see][]{Li2010} to the typical 
energy released in the explosion, $E_{\rm SN}=10^{51}$ erg, to the the number of SN per solar mass, 
$\eta_{\rm SN}=8\times 10^{-3}~M_{\odot}^{-1}$ (for a Chabrier IMF), to the efficiency energy is able to reheat the 
disk gas $\epsilon_{\rm halo}=0.35$ \citep{Li2010} and to $\Delta M_{\ast}$, the mass converted in stars. The gas 
reheating produced by energy injection is proportional to the stars formed 
$\Delta M_{\rm reheat} = \epsilon_{\rm disk}\Delta M_{\ast}$, where $\epsilon_{\rm disk}=3.5$ \citep{Li2010}. 
The hot gas ejected is proportional to $\Delta E_{\rm SN}-\Delta E_{\rm hot}$, where 
$\Delta E_{\rm hot}= 0.5\Delta M_{\rm reheat} \eta_{\rm SN}E_{\rm SN}$ is the thermal energy change produced by the 
reheated gas \cite[see][]{Li2010}.
}
{
Reionization reduces the baryon content and takes place in the redshift range 11.5-15. 
}

In the following stage, SN explosions {repeatedly} eject gas, {and thus lower stellar density}. Feedback 
destroys the smallest clumps soon after a small part of their mass is transformed into stars \citep{Nipoti2015}.

For masses $M\simeq 6 \times 10^{11} M_{\odot}$, AGN quenching must be taken into account \citep{Cattaneo2006}. AGN 
feedback {follows the prescription of} \cite{Martizzi2012,Martizzi2012a}. A Super-Massive-Black-Hole (SMBH) is 
created when the star {density exceeds $2.4 \times 10^6 M_{\odot}/{\rm kpc}^3$, the gas density reaches 10 times 
the stellar density, and the 3D velocity dispersion exceeds 100 $km/s$. Each initial (seed) black hole mass starts at 
$10^5~M_{\odot}$. Mass accretion into the SMBH and AGN feedback were implemented modifying the model by 
\cite{Booth2009} as in \cite{Martizzi2012}.}

{
Our model {\it does not} contain data fitted parameters. Its parameters are the same found in simulations (e.g., star 
formation rate, density treshold of star formation, and other parameters previously discussed and also found in the 
references given).
}

%
%

{
The model demonstrated its robustness predicting before simulations the density flattening and shape produced by 
heating of DM, for galaxies \cite{DelPopolo2009,DelPopolo2010,DelPopolo2012a} in agreement with {subsequent} SPH 
simulations \citep[e.g.][]{Governato2010,Governato2012} 
\cite[see also Fig.~4 of][for a direct comparison]{DelPopolo2011}, and similarly for clusters of galaxies  
\cite{DelPopolo2012b} in agreement with hydro-simulations of \citep{Martizzi2013}, the inner density slope dependence 
on the halo mass and on the total baryonic content to the total mass ratio \cite{DelPopolo2010,DelPopolo2016a}, in 
agreement with \cite{DiCintio2014}. The slope was shown by \cite{DelPopolo2010,DelPopolo2016a} to depend also on the 
angular momentum. In \cite{DelPopolo2016b}, the stellar and baryonic Tully-Fisher, Faber-Jackson and the 
stellar mass vs. halo mass (SMH) relations were shown in agreement with simulations.

Finally, the correct DM profile inner slope dependence on halo mass was explained over 6 order of magnitudes in halo 
mass, from dwarves to clusters\cite{DelPopolo2009,DelPopolo2010,DelPopolo2012a,DelPopolo2012b}, a range no other 
model achieved.}

We wrote an Appendix with more details of the model, in the final part of the paper.

\section{Diversity}\label{sect:diversity}

This section presents the results of our model (within $\Lambda$CDM cosmology), simulating 100 galaxies, likened to the nearby galaxies with high-quality RCs of the \cite{Oman2015} sample. Such simulations are made with stellar masses in the range $M_{\ast}=6 \times 10^6-10^{11}~M_{\odot}$. 

%
%

The model produces a fair sample of galaxies, with a mass distribution following the \cite{DelPopolo2017} halo mass function. As in \cite[Eq.~12]{Shen2003}, their size distribution is log-normal, derived from the corresponding spin parameter log-normal distribution. The RCs of groups of simulated galaxies are compared with those of similar observed galaxies (e.g. comparable halo and baryonic mass). 

Although average dwarf galaxies RCs are cored, the range of RCs fits covers inner slopes from $\alpha \simeq 0$ to cuspy values. Even cored RCs display up to a factor 10 variations among galaxies from similar haloes \cite{KuziodeNaray2010}, and that variability increases for higher masses. Many haloes were found to contain similar dwarfs with either cored or cuspy profiles \cite{Simon2005}, while other studies \cite{deBlok2008} reported a trend for less massive galaxies to display flatter profiles.

This apparent diversity was constrained in \cite{Oman2015}, confronting $V_{\rm 2kpc}$, the 2~kpc radius circular velocity, to $V_{\rm max}$, a fixed maximum circular velocity. The scatter in $V_{\rm 2kpc}$ is small for the range $50~{\rm km/s} <V_{\rm max} <250~{\rm km/s}$. A study based on the SIDM \cite{Creasey2017} showed the diversity solution requires additional baryonic physics, and thus that the SIDM alone is unable to explain the scatter.

\begin{figure*}
\includegraphics[height=8cm,angle=0]{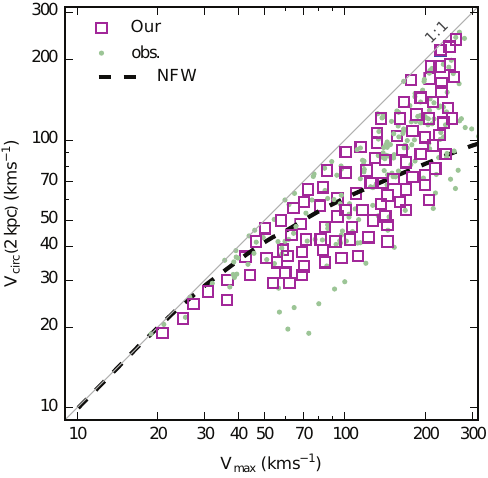}
\includegraphics[width=9.5cm,height=8cm,angle=0]{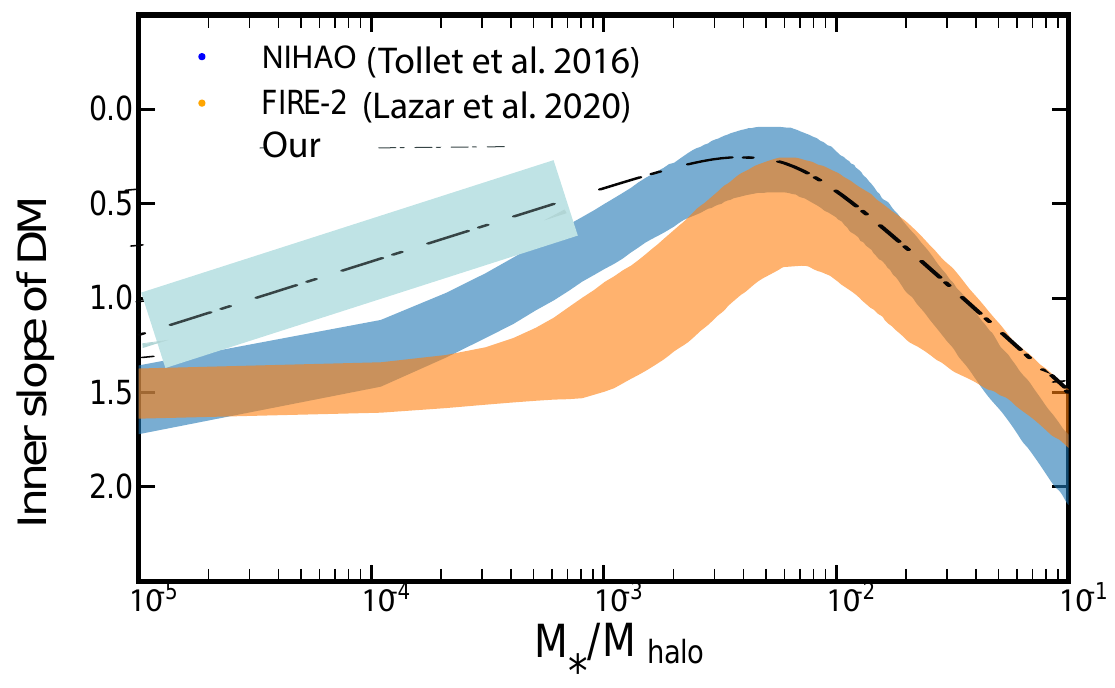}
 \caption[justified]{
 Left: $V_{\rm 2kpc}$-$V_{\rm max}$ correlation under effects of baryonic physics. If all haloes followed the NFW profile, the dashed line would be their correlation. Dots mark galaxies from \cite{Oman2015}, while squares indicate our model's predictions.
\\
 Right: DM haloes inner slope vs. $M_{\star}/M_{\rm halo}$. Results from the for the models of \citep{Lazar2020}, \citep{Tollet2016} and the present paper's are represented by shaded regions, respectively, in orange, blue and light blue. Note the light blue region is not represented throughout so as not to clutter the plot. Instead, our model is presented through the whole graph by the dot-dashed line.
 }
\label{fig:Lelli}
\end{figure*}

This confrontation of $V_{\rm 2kpc}$ with $V_{\rm max}$, the outermost, thus maximum, observed circular velocity, is shown in the left panel of Fig.~\ref{fig:Lelli}. We added there the expected correlation if all haloes followed a NFW density profile. Observations from \cite{Oman2015} are marked with dots, while open squares represent our model's predictions. 

The stellar masses of the \cite{Oman2015} sample range in $M_{\star}\simeq 5.10^6-10^{11.5}~M_{\odot}$, which translates into circular velocities within $15-300~{\rm km/s}$. Their comparison with our model was made possible choosing \cite{Lelli2016}'s stellar mass range an computing their $V_{\rm 2kpc}$ and $V_{\rm max}$.

From our model's results, scatter in $V_{\rm 2kpc}$, at a given $V_{\rm max}$, can reach up to a factor four. The $\Lambda$CDM model for haloes, with restricted freedom to one parameter (concentration or halo mass), with  only provides cuspy, self-similar halo profiles. It thus cannot explain this scatter or the cored dwarfs, because the larger DM/baryon ratios in cusps induce a ''freeze`` in the $V_{\rm 2kpc}$ scatter that originates in the baryon distribution. Our model allows that scatter because baryon physics, by heating and enlarging galaxies, decreases the radio of inner DM, and induces a mass dependent profile, as depicted in \cite{DelPopolo2010,DelPopolo2016b}. In this way, the tendencies for cuspy profiles of MW types, for cored profiles of dwarfs, and for intermediate behaviour in Ultra-faint dwarfs, produce the scatter in Fig.~\ref{fig:Lelli}, that originates in the discussed \cite{DelPopolo2012a} environment and mass dependence effects on core formation.

It is therefore no surprise that our model succeeds in predicting the observed RCs' scatter and distribution, since the haloes of simulated galaxies including baryon physics react according to their environment. The 1-$\sigma$ DM halo's scatters 
are represented in Fig.~\ref{fig:Lelli}'s right panel in shaded regions as a function of stars to halo mass ratio: in  blue, for the results of \citep{Tollet2016}, in orange from \citep{Lazar2020}, and in light blue for the present paper's results, supplemented by our inner slopes as a dot-dashed line. Note the shaded light blue is only shown partially so as not to clutter the plot. The curves for these results were built from power law fits to the inner DM profiles, in the fashion of \citep{DiCintio2014,Tollet2016}.

From Fig.~\ref{fig:Lelli}, it is clear that the core formation mechanism, and the resulting inner slope $\alpha$, strongly depend on $M_{\ast}/M_{\rm halo}$.\footnote{The $\alpha$ vs. $M_{\ast}/M_{\rm halo}$ correlation can be recast, with the help of the \cite{Moster2013} relation, as $\alpha$ vs. $M_{\ast}$.} The maximum baryionic effect corresponds to the minimum $\alpha$ value at $M_{\ast}/M_{\rm halo} \simeq 10^{-2}$, i.e. $M_{\ast} \simeq 10^8 M_{\odot}$ \citep{DiCintio2014,Tollet2016,DelPopolo2016a}. Below this threshold, the relative reduction in stars ($M_{\ast}/M_{\rm halo}$) steepens the profiles. Since $V_{\rm max}\propto M_{\ast}$ and the profile steepens for $M_{\ast}\lesssim 10^8 M_{\odot}$, a self-similar RCs formation, in the same way as for the NFW profile, is expected. Indeed, Fig.~\ref{fig:Lelli}'s left panel displays such behaviour. Above the threashold, $M_{\ast}\geq 10^8 M_{\odot}$, the increased stellar mass deepens the central potential well, thus reducing the effect of baryons and, again, steepening the inner profile. In addition to the simple potential well reinforcement, AGN feedback also counters baryon cooling by modifying star formation. Accounting for all this for $V_{\rm max}\ge 150~{\rm km/s}$, our model thus agrees with the \cite{Oman2015} sample. While usual hydrodynamical simulations focus on isolated galaxies, tidal interactions included in our model give it more environment dependence.

\section{Outliers: Dwarf and LSBs galaxies}\label{sect:dwarf}

%

%

In the previous section, we presented a sample constituted by 100 galaxies generated by means of our model with {similar characteristics} to the \cite{Oman2015} sample. 

In the sample there are extreme outliers which cannot explained  within the $\Lambda$CDM
scenario. Two peculiar cases highlighted in \cite{Oman2015} (but also present in the SPARC smple), are UGC~5721 and IC~2574. The two galaxies have similar circular velocity
measured at the outermost point of the rotation curve, $80 \rm km/s$, meaning that  they populate dark matter halos with similar mass. However, the shapes of the inner
regions are remarkably different: whereas UGC~5721 is consistent with a `cuspy' NFW profile, IC~2574 has a very extended core. 

\begin{figure}
\includegraphics[height=5cm,angle=0]{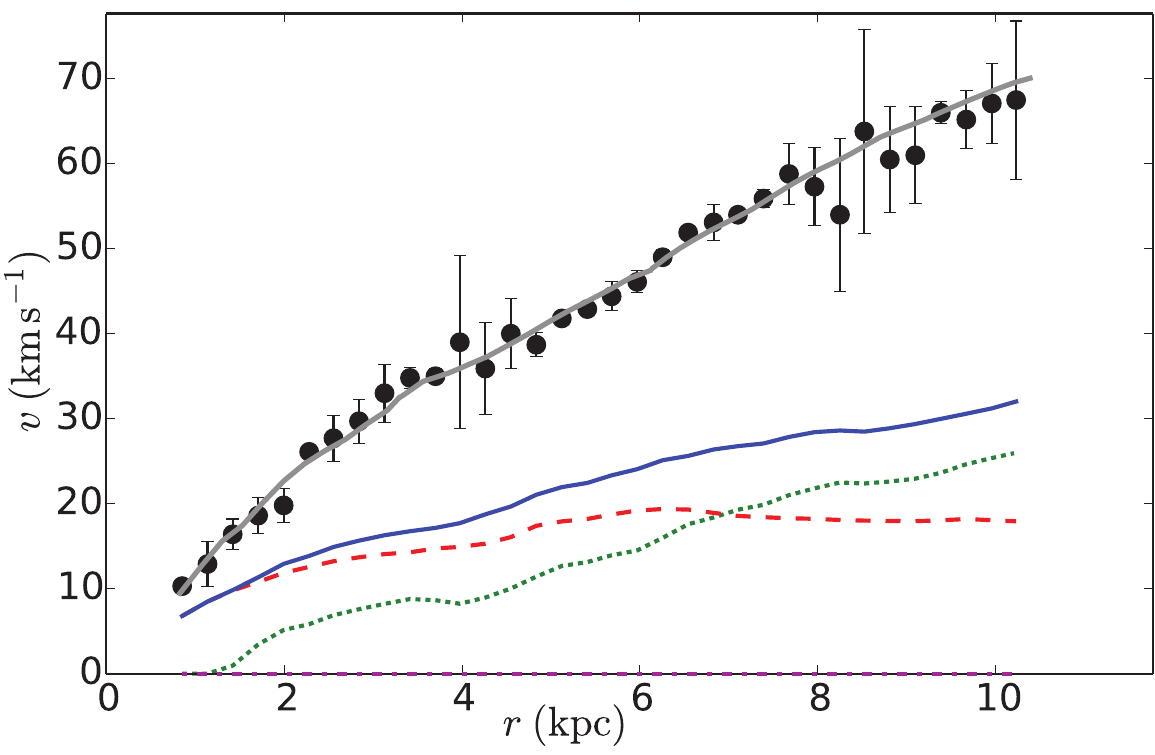}
 \caption[justified]{
 IC~2574 rotation curves. Observations from the SPARC catalogue are marked as dots with error bars, while our model's RC is shown as solid grey line. Baryonic contributions to the RC are given for stars, as short dashed 
 green line, the gas disk, as long dashed red line, and their sum, the total baryonic mass, as solid blue line. The final RC should also be compared with the SIDM based model from \cite{Kaplinghat2016}.
 }
 \label{fig:RC}
\end{figure}

As IC~2574 appears on the distribution's edge, particular attention was given to it: a re-simulation in that rarefied range was proceeded, aiming at outputting a galaxy within a few percents of its mass profile. No further adjustments was taken than this focus on the low probability edge. The resulting RC is presented in Fig.~\ref{fig:RC}, compared with IC~2574's RC and with its SIDM model from \cite{Kaplinghat2016}.

Repeated simulations were necessary to obtain DM halo and baryonic masses\footnote{Including the resulting density profile, and thus RC, as well.} within 10\% of the observed IC~2574's. Fig.~\ref{fig:RC}'s RCs display IC~2574's data as dots with error bars, and the model prediction as solid grey line, with each contributions from the total baryonic mass, decomposed into gas and stars, respectively as solid blue, long dashed red and short dashed 
green lines.

Contrary to the claims that IC~2574 was a stumbling block for core formation scenarii \cite{Oman2015}, Fig.~\ref{fig:RC} demonstrates the success of our model overall in explaining IC~2574, as well as in its baryonic components when compared with \citep{Blais-Ouellette2001} or their counterparts in the SPARC mass models \cite{Oh2008}. Note that the discrepant baryonic contributions (gas, stars) between \cite{Kaplinghat2016}, SPARC and \citep{Blais-Ouellette2001} originate from their different DM properties. Our simulated host galaxy halo carries $1.8 \times 10^{10}~M_{\odot}$, with $M_{\ast}=10^{8.7} 
M_{\odot}$ in stellar mass . The virial mass was derived from the stellar mass following \cite{Moster2013}, yielding a galactic effective radius $R_{\rm eff}=2.8~{\rm kpc}$.

A recent SN feedback based model \cite{Santos2017} reobtained IC~2574's RC, with an approach similar to our's.

\begin{figure}
 \includegraphics[height=7cm,angle=0]{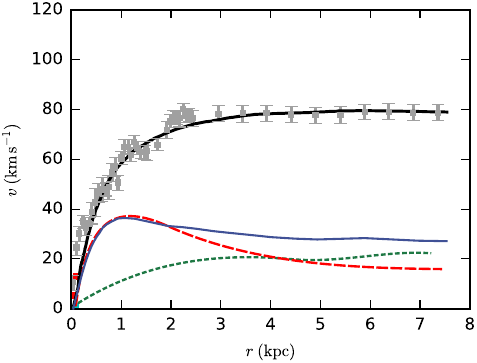}
 \caption[justified]{
 UGC~5721 rotation curves. Observations are marked as dots with error bars, while our model's RC is shown as solid black line. Baryonic contributions to the RC are given for stars, as long dashed red line, the gas disk, as short dashed green line, and their sum, the total baryonic mass, as solid blue line.
 }
 \label{fig:UGC5721}
\end{figure}

Encouraged by this success, we applied the same method to UGC~5721, which results are shown in Fig.~\ref{fig:UGC5721}. It is known that all its components taken into account yield a steeper cuspy profile circular velocity than its expected NFW halo.

As in the IC~2574 case, repeated simulations were conducted until convergence within 10\% of UGC~5721's observed halo and baryonic masse.

Fig.~\ref{fig:UGC5721} represents again UGC~5721's RC data with dots and their error bars, while the model's RC prediction is this time a thick black line. The baryonic RC contributions are now depicted, 
the total baryonic mass as solid blue. It is also 
decomposed into gas and stars, respectively\ displayed 
as long dashed 
red and short dashed 
green 
lines. From Fig.~\ref{fig:UGC5721} can be concluded that our simulated galaxy describes very well UGC~5721's RCs, overall and in baryonic components.


\section{Conclusions}\label{sect:conclusions}
At small scales, issues appear in the predictions of the $\Lambda$CDM, ''concordance``, model, in particular with the excess in predictions of DM content for galaxies and clusters inner regions. However, accounting for the effects of baryons, this paper showed a solution to that issue, that incidentally unifies as well the solutions of other small scale problems, within the $\Lambda$CDM paradigm, thus without any need for modifying the nature of DM \citep[i.e., contrary to the][SIDM-based solution]{Kaplinghat2016}. That solution, a dynamical friction centered semi-analytic model, outputted a sample of simulated galaxies and clusters resembling the \cite{Oman2015} set, with some galaxies akin to those scrutinised in \citep{Kaplinghat2016}. That sample was the base of a ``diversity" problem exploration, comparing, with the \cite{Oman2015} compilation, the galaxies' $V_{\rm 2kpc}$ at fixed $V_{\rm max}$. It also showed the variety of RCs produced by baryon physics, in particular their environment, as well as total and stellar mass dependence, which explains the $V_{\rm 2kpc}$-$V_{\rm max}$ plane observed scatter. Fig.~\ref{fig:Lelli} showed how including baryon physics does explain well the galaxies RCs' diversity. However, some ouliers appear, with RCs deemed not to fit within the $\Lambda$CDM paradigm (e.g., IC~2574). A refocus of our model was again able to produce good agreement with observations for the case of UGC~5721, and even for the dreaded case of IC~2574 (Fig.~\ref{fig:RC}).

\section*{Acknowledgments}

MLeD acknowledges the financial support by the Lanzhou University starting
fund, the Fundamental Research Funds for the Central Universities
(Grant No. lzujbky-2019-25), National Science Foundation of China (grant No. 12047501) and the 111 Project under Grant No. B20063. 
The authors wish to thank Maksym Deliyergiyev for some calculations.

\newpage

\appendix

\section{Theoretical Model}
\label{sec:Implementation}

This section recalls the model employed in this work. 
The spherical collapse models \citep{Gunn1972,Bertschinger1985,Hoffman1985,Ryden1987,Ascasibar2004,Williams2004} was very significant improved in \cite[e.g.][]{DelPopolo2009,DelPopolo2009a} to include the effects of
\begin{enumerate}[leftmargin=*,label=$-$,noitemsep,partopsep=0pt,topsep=0pt,parsep=0pt]
 \item random angular momentum induced by random motion during the collapse phase of haloes \citep[e.g.,][]{Ryden1987,Williams2004},
 \item ordered angular momentum induced by tidal torques \citep[e.g.,][]{Ryden1988,DelPopolo1997,DelPopolo2000}, 
\end{enumerate}
and was furthered to include the consequences of
\begin{enumerate}[resume,leftmargin=*,label=$-$,noitemsep,partopsep=0pt,topsep=0pt,parsep=0pt]
 \item adiabatic contraction \citep[e.g.,][]{Blumenthal1986,Gnedin2004, Klypin2002,Gustafsson2006}, 
 \item dynamical friction between DM and baryonic gas and stellar clumps 
\citep{ElZant2001,ElZant2004,Ma2004,RomanoDiaz2008,RomanoDiaz2009,DelPopolo2009,Cole2011,Inoue2011, Nipoti2015}, 
 \item gas cooling, star formation, photoionization, supernova, and AGN feedback \citep{DeLucia2008,Li2010,Martizzi2012} and 
 \item DE \citep{DelPopolo2013a,DelPopolo2013b,DelPopolo2013c}, 
\end{enumerate}
and was further refined in  \cite{DelPopolo2014a,DelPopolo2016a,DelPopolo2016b,DelPopolo:2016skd}. This model produced results on 
\begin{enumerate}[leftmargin=*,label=$\centerdot$,noitemsep,partopsep=0pt,topsep=0pt,parsep=0pt]
                                                                                                                    \item the universality of density profiles \citep{DelPopolo2010,DelPopolo2011}, 
                                                                                                                \item specific features of density profiles in 
                                            \begin{enumerate}[leftmargin=*,label=$\cdot$,noitemsep,partopsep=0pt,topsep=0pt,parsep=0pt]
                                                                                                                                                                                    \item galaxies \citep{DelPopolo2012a,DelPopolo2014} and
                                                                                                                                                                                    \item clusters \citep{DelPopolo2012b,DelPopolo2014},  
                                                                                                                                                            \end{enumerate}
                                                                                                                                                            \end{enumerate}
as well as a focus on                                                                                                                                                
\begin{enumerate}[resume,leftmargin=*,label=$\centerdot$,noitemsep,partopsep=0pt,topsep=0pt,parsep=0pt]
                                                                                                                                                        
\item galaxies inner surface-density \citep*{DelPopolo2013d}.
                                                                                                                                        \end{enumerate}

Although the model's key mechanism resides in dynamical friction (DFBC), we stress out that it includes all of the above effects (including SNF) that each only contribute at the level of some \%.

Its implementation occurs in several stages:
\begin{enumerate}
 \item The diffuse proto-structure of gas and DM expands, in the linear phase, to a maximum radius before DM re-collapses into a potential well, where baryons will fall.
\item In their radiative clumping, baryons form stars at the halo centre.
\item Then four effects happen in parallel
\begin{enumerate}
 \item the DM central cusp increases from baryons adiabatic contraction (at $z \simeq 5$ in the case of $10^9 M_{\odot}$ galaxies \citep{DelPopolo2009})
 \item the galactic centre also receive clumps that collapse from baryons-DM dynamical friction (DF)
 \item the DF energy and angular momentum (AM) transfer to DM  \citep[and stars][]{Read2005,Pontzen2012,Teyssier2013} results in an opposite effect to adiabatic contraction, and reduces the halo central density \citep{ElZant2001,ElZant2004}.
 \item the balance between adiabatic contraction and DF can result in heating cusps and forming cores, i.e. in dwarf spheroidals and spirals, while the deeper potential wells of giant galaxies keeps their profile steeper.
\end{enumerate}
\item The effect of DF 
adds to that of tidal torques (ordered AM), and random AM. 
  \item \label{SNmech} Finally, the core further slightly (few percent) enlarges from the decrease of stellar density due to successive gas expulsion from supernovae explosions, and from the disruption of the smallest gas clumps, once they have partially converted to stars \citep[see][]{Nipoti2015}.
\end{enumerate}

\subsection{Model treatment of density profile}

Starting from a Hubble expansion, the spherical model of density perturbations expands linearly until reaching a turn-around maximum and reverting into collapse \citep{Gunn1977,Fillmore1984}. A Lagrange particle approach yields the final density profile
\begin{equation}\label{eq:dturnnn}
 \rho(x)=\frac{\rho_{\rm ta}(x_{\rm m})}{f(x_{\rm i})^3}
 \left[1+\frac{d\ln{f(x_{\rm i})}}{d\ln{g(x_{\rm i})}}\right]^{-1}\;,
\end{equation}
with initial and turn-around radius, ,resp. $x_{\rm i}$ and $x_{\rm m}(x_{\rm i})$, collapse factor $f(x_{\rm i})=x/x_{\rm m}(x_{\rm i})$, and turnaround density $\rho_{\rm ta}(x_{\rm m})$. The turn-around radius is obtained with
\begin{equation}
 x_{\rm m}=g(x_{\rm i})=x_{\rm i}\frac{1+\overline{\delta}_{\rm i}}
 {\overline{\delta}_{\rm i}-(\Omega_{\rm i}^{-1}-1)}\;, 
\end{equation}
where we used $\Omega_{\rm i}$ for the density parameter, and $\overline{\delta}_{\rm i}$ for the average 
overdensity inside a DM and baryons shell.

The model starts with all baryons in gas form with $f_{\rm b}=0.17\pm 0.01$ for the ''universal baryon fraction`` \citep{Komatsu2009} \citep[set to 0.167 in][]{Komatsu2011}, before star formation proceeds as described below.

Tidal torque theory (TTT) allows to compute the ''specific ordered angular momentum``, $h$, exerted on smaller scales from larger scales tidal torques  \citep{Hoyle1953,Peebles1969,White1984,Ryden1988,Eisenstein1995}, while the ''random angular momentum``, $j$, 
{is related to }orbits eccentricity $e=\left(\frac{r_{\rm min}}{r_{\rm max}}\right)$ \citep{AvilaReese1998}, obtained from the apocentric radius $r_{\rm max}$, the pericentric radius$r_{\rm min}$ and corrected from the system's dynamical state effects advocated by \cite{Ascasibar2004}, using the spherically averaged turnaround radius $r_{\rm ta}=x_{\rm m}(x_{\rm i})$ and the maximum radius of the halo $r_{\rm max}<0.1 r_{\rm ta}$
\begin{equation}
e(r_{\rm max})\simeq 0.8\left(\frac{r_{\rm max}}{r_{\rm ta}}\right)^{0.1}\;.
\end{equation}

These corrections to the density profile are compounded also with its steepening from the adiabatic compression following \cite{Gnedin2004} and the effect of DF introduced in the equation of motion by a DF force \citep[see][Eq. A14]{DelPopolo2009}.

\subsection{Effects of baryons, discs, and clumps}

The baryon gas halo settles into a stable, rotationally supported, disk, in the case of spiral galaxies. Their size and mass result from solving the equation of motion, and lead to a solution of the angular momentum catastrophe (AMC) \citep[Section 3.2, Fig. 3, and 4 of][]{DelPopolo2014}, obtaining realistic disc size and mass.

Notwithstanding stabilization from the shear force, Jean's criterion shows the appearance of instability for denser discs. The condition for this appearance and subsequent clump formation was found by Toomre \cite{Toomre1964}, involving the 1-D velocity dispersion $\sigma$,\footnote{$\simeq 20-80$ km/s in most clump hosting galaxies} angular velocity $\Omega$, surface density $\Sigma$, related to the adiabatic sound speed $c_s$, and the epicyclic frequency $\kappa$
\begin{equation}
Q \simeq \sigma \Omega/(\pi G \Sigma)=\frac{c_s \kappa}{\pi G \Sigma}<1\;.
\end{equation}
The solution to the perturbation dispersion relation $d \omega^2/d k=0$ for $Q<1$ yields the fastest growing mode  $k_{\rm inst}=\frac{\pi G \Sigma}{c_s^2}$ (see \cite{BinneyTremaine1987} or \cite[Eq. 6]{Nipoti2015}). That condition allows to compute the clumps radii in galaxies \citep{Krumholz2010}
\begin{equation}
 R \simeq 7 G \Sigma/\Omega^2 \simeq 1 {\rm kpc}\;.
\end{equation}
Marginally unstable discs ($Q \simeq 1$) with maximal velocity dispersion have a total mass three times larger than that of the cold disc, and form clumps 
$\simeq 10$ \% of their disk mass $M_d$ \citep{Dekel2009}.

Objects of masses few times $10^{10}~M_{\odot}$, found in $5 \times 10^{11} M_{\odot}$ haloes at $z \simeq 2$, are in a marginally unstable phase for $\simeq 1$~Gyr. Generally the main properties of clumps are similar to those found by \cite{Ceverino2012}.

In agreement with 
\cite{Ma2004,Nipoti2004,RomanoDiaz2008,RomanoDiaz2009,DelPopolo2009,Cole2011,Inoue2011,DelPopolo2014d,Nipoti2015}, 
energy and AM transfer from clumps to DM flatten the profile more efficiently in smaller haloes.

%

\subsubsection{Computing the clumps life-time}

Evidence for existence of the clumps produced by the model can be traced both in simulations \citep[e.g.,][]
{Ceverino2010,Perez2013,Perret2013,Ceverino2013,Ceverino2014,Bournaud2014,Behrendt2016}, and observations. High redshift galaxies have been found to contain clump clusters or clumpy structures that leads to call them chain galaxies \citep[e.g.,][]{Elmegreen2004,Elmegreen2009,Genzel2011}. The HST Ultra Deep Field encompasses galaxies with massive star-forming clumps \citep{Guo2012,Wuyts2013}, many at $z=1-3$ \citep{Guo2015}, some in deeper fields $z \leq 6$ \cite{Elmegreen2007}.

Such clumpy structures are expected to originate from self-gravity instability in very gas-rich disc, from radiative cooling in the acreting dense gas
\citep[e.g.,][]{Noguchi1998,Noguchi1999,Aumer2010,Ceverino2010,Ceverino2012}. Their effect on halo central density depend crucially on the clump lifetime: should their disruption through stellar feedback still allow them sufficient time
to sink to the galaxy centre, they can turn a cusp into a core. A clump's ability to form a bound stellar system is assessed through its stellar feedback mass fraction loss,
$e_f$, and its formed stars
mass fraction, $\varepsilon=1-e_f$. Simulations and analytical models agree that most of the mass of
such 
group of stars will remain bound for $\epsilon \geq 0.5$ \cite{Baumgardt2007}. The radiation feedback efficiency can be estimated, using 
 a. the dimensionless star-formation rate efficiency $\epsilon_{eff}=\frac{\dot{M_*}}{M/t_{ff}}$. This 
is simply the ratio between 
free-fall time, $t_{ff}$, and the 
depletion time for a stellar mass $M_*$. In 
its reduced version it reads $\epsilon_{eff},_{-2}=\epsilon_{eff}/0.01$,\\
b. the reduced dimensionless surface density $\Sigma_1=\frac{\Sigma}{0.1 g/cm^2}$, and

\vspace{-.4cm}\noindent to obtain the expulsion fraction $e_f=1-\varepsilon=0.086 (\Sigma_1 M_9)^{-1/4} \epsilon_{eff},_{-2}$ \citep{Krumholz2010}. Ref.~\citep{Krumholz2007} estimated, for a large sample of environments, densities, size and scales, that $\epsilon_{eff}\simeq 0.01$. Furthermore, $e_f=0.15$ and $\varepsilon=0.85$ for typical clumps with masses  $M\simeq10^9 M_{\odot}$. Therefore, the clump mass loss before they reach the centre of the galactic halo should be small. However, such conclusion and the expulsion fraction method are valid for smaller, more compact clumps in smaller galaxies. Such context only produces clumps that survive all the way to the centre.

Alternately, comparing a clump lifetime to its migration time to the centre, one can also obtain clump disruption. Migration time is the result of DF and TTT: for a $10^9 M_{\odot}$ clump, it yields $\simeq 200$ Myrs \citep[see Eq. 1 of][Eq. 18]{Genzel2011,Nipoti2015}. Coincidents 
expansion 
and migration timescales were 
computed from the  Sedov-Taylor solution \citep[Eqs. 8,9]{Genzel2011}.

Clump lifetime has been much studied. Ceverino \emph{et al.}, finding clumps in Jean's equilibrium and rotational support, from hydrodynamical simulations \citep{Ceverino2010}, construed their
long lifetime ($\simeq 2 \times 10^8$ Myr). This agrees with several approaches: in local systems forming stars and coinciding with the Kennicutt-Schmidt law,  \citep{Krumholz2010} found such lifetimes. This is because as clumps retained gas, and formed bound star groups, they 
had time to migrate to the galactic centre. Simulations from \cite{Elmegreen2008} confirmed it. Other simulations with proper account of stellar feedback, e.g. non-thermal and radiative feedback mechanisms, 
also obtained long-lived clumps reaching galactic centre \citep[SNF, radiation pressure, etc][]{Perret2013,Bournaud2014,Ceverino2013}. Finally, 
the same was obtained with any reasonable amount of feedback \cite{Perez2013}. The expansion, gas expulsion, and metal enrichment, time scales (respectively $>100$ Myrs, 170-1600 Myrs, and $\simeq 200$ Myrs) obtained by \cite{Genzel2011} to estimate clump ages also bring strong evidence for long-lived clumps. Lastly, comparison between similar low and high redshift clumps observations \citep[in radius, mass,][]{Elmegreen2013,Garland2015,Mandelker2015} supports clump stability.

\subsection{Model treatment of feedback and star formation}

Star formation, reionisation, gas cooling, and SNF in the model are built along \citep[Secs.~2.2.2 and~2.2.3]{DeLucia2008,Li2010}.

\begin{description}
 \item[Reionisation] acts for $z=11.5-15$ by decreasing the baryon fraction as
\begin{equation}
f_{\rm b, halo}(z,M_{\rm vir})=\frac{f_{\rm b}}{[1+0.26 M_{\rm F}(z)/M_{\rm vir}]^3}\;,
\end{equation}
\citep{Li2010}, using the virial mass,$M_{\rm vir}$, and the ``filtering mass'' \citep[see][]{Kravtsov2004}, $M_{\rm F}$. 
\item[Gas cooling] follows from a cooling flow model \citep[e.g.,][see Sect. 2.2.2]{White1991,Li2010}.
\item[Star formation] arises from gas conversion into stars when it has settled in a disk. The gas mass conversion into stars during a given time interval $\Delta t$, which we take as the disc dynamical time $t_{\rm dyn}$, is given by
\begin{equation}
 \Delta M_{\ast}=\psi\Delta t\;,
\end{equation}
where the star formation rate $\psi$ comes from the gas mass above the density threshold $n>9.3/{\rm cm^3}$ \citep[fixed as in][]{DiCintio2014} according to \citep[see][for more details]{DeLucia2008}
\begin{equation}
\psi=0.03 M_{\rm sf}/t_{\rm dyn}\;.
\end{equation}
\item[SNF] follows \cite{Croton2006}, where SN explosions inject energy in the system. This energy can be 
calculated from a Chabrier IMF \cite{Chabrier2003}, using \begin{itemize}[leftmargin=*,label=$-$,noitemsep,partopsep=0pt,topsep=0pt,parsep=0pt]
                                                           \item the disc gas reheating energy efficiency $\epsilon_{\rm halo}$,
                                                           \item the available star mass $\Delta M_{\ast}$, 
                                                           \item that mass conversion into SN measured with the SN number per solar mass as $\eta_{\rm SN}=8\times 10^{-3}/M_{\odot}$, and
                                                           \item the typical energy an SN explosion releases $E_{\rm SN}=10^{51}$ erg,
                                                          \end{itemize}
 to obtain
\begin{equation}
 \Delta E_{\rm SN}=0.5\epsilon_{\rm halo}\Delta M_{\ast} \eta_{\rm SN}E_{\rm SN}\;.
\end{equation}
This released energy from SNs into the hot halo gas in the form of reheated disk gas then compares with the reheating energy $\Delta E_{\rm hot}$ which that same amount of gas should acquire if its injection in the halo should keep its specific energy constant, that is if the new gas would remain at equilibrium with the halo hot gas. That 
amount of disk gas the SN and stars radiation have reheated, $\Delta M_{\rm reheat}$, 
 since it is produced from stars radiations, is 
proportional to their mass
\begin{equation}
 \Delta M_{\rm reheat} = 3.5 \Delta M_{\ast}\;.
\end{equation}
Since
the halo hot gas specific energy 
corresponds to the Virial equilibrium specific kinetic energy $\frac{V^2_{\rm vir}}{2}$, keeping this energy constant under addition of that 
reheated gas leads to define the equilibrium reheating energy as
\begin{equation}
\Delta E_{\rm hot}= 0.5\Delta M_{\rm reheat} 
V^2_{\rm vir}\;.
\end{equation}
The comparison with the actual energy of the gas injected from the disk into the halo by SNs 
gives the threashold ($\Delta E_{\rm SN}>\Delta E_{\rm hot}$) beyond which gas is expelled, 
the available energy to expel the reheated gas, and thus the amount of gas ejected from that extra energy
\begin{equation}
 \Delta M_{\rm eject}=\frac{\Delta E_{\rm SN}-\Delta E_{\rm hot}}{0.5 V^2_{\rm vir}}\;.
\end{equation}\\
Contrary to SNF based models such as \cite{DiCintio2014}, our mechanism for cusp flattening initiates before the star formation epoch. Since it uses 
a gravitational energy source, it 
is thus less limited in available time and energy. Only after DF shapes the core can Stellar and SN feedback occurs, which then disrupt gas clouds in the core \citep[similarly to][]{Nipoti2015}.
\item[AGN feedback] occurs when a central Super-Massive-Black-Hole (SMBH) is formed. We follow the prescriptions of \cite{Martizzi2012,Martizzi2012a}, modifying the \cite{Booth2009} model for SMBH mass accretion and AGN feedback: a seed $10^5~M_{\odot}$ SMBH forms when stellar density, reduced gas density ($\rho_{gas}/10$) and 3D velocity dispersion exceed the thresholds $2.4 \times 10^6 M_{\odot}/{\rm kpc}^3$ and 100 $km/s$, which then accretes. Significant AGN quenching starts above $M\simeq 6 \times 10^{11} M_{\odot}$ \citep{Cattaneo2006}.
\end{description}

\subsection{Model robustness}

We point out that the model demonstrated its robustness in various behaviours:

a. The cusp flattening from DM heating by collapsing baryonic clumps predicted for galaxies and clusters is in agreement with following studies 
        \citep{ElZant2001,ElZant2004,RomanoDiaz2008,RomanoDiaz2009,Cole2011,Inoue2011,Nipoti2015}. A comparison with \cite{Governato2010}'s  SPH simulations was made in \citep[][Fig. 4]{DelPopolo2011}.
        b. it aforetime predicted the correct shape of galaxies density profiles  \citep{DelPopolo2009,DelPopolo2009a}, ahead of SPH simulations of \cite{Governato2010,Governato2012}, and of clusters density profiles \citep{DelPopolo2012b}  anteriorly of \cite{Martizzi2013}. \footnote{Note that \cite{Governato2010,Governato2012} and \cite{Martizzi2013} adopted different dominant mecanisms.}
c. it aforetime predicted the halo mass dependence of cusps inner slope \cite[Fig. 2a solid line]{DelPopolo2010} beforehand the similar result in the non-extrapolated part of the plot in \cite[Fig. 6]{DiCintio2014}, expressed in terms of $V_c$, as it corresponds to $2.8 \times 10^{-2} M_{\rm vir}^{0.316}$ \citep{Klypin2011}.
d. it also preceded \cite{DiCintio2014} in predicting \cite[see][]{DelPopolo2012b} that the inner slope depends on the 
        total baryonic content to total mass ratio.
e. it compares well with simulations \citep{DiCintio2014} on the inner slope change with mass \citep[Fig. 1]{DelPopolo2016a,DelPopolo2016b}.
f. it moreover provides a comparison of the Tully-Fisher and Faber-Jackson, $M_{Star}-M_{halo}$, relationships with simulations \citep[Figs. 4, 5]{DelPopolo2016a,DelPopolo2016b}.

\bibliographystyle{apsrev4-1}
\bibliography{old_MasterBib1.2,old_MasterBib2,biblioNS}

\label{lastpage}

\end{document}